# An Investigation of Performance versus Security in Cognitive Radio Networks with Supporting Cloud Platforms

Kurniawan D. Irianto, Demetres D. Kouvatsos

*Abstract*—The growth of wireless devices affects the availability of limited frequencies or spectrum bands as it has been known that spectrum bands are a natural resource that cannot be added. Meanwhile, the licensed frequencies are idle most of the time. Cognitive radio is one of the solutions to solve those problems. Cognitive radio is a promising technology that allows the unlicensed users known as secondary users (SUs) to access licensed bands without making interference to licensed users or primary users (PUs). As cloud computing has become popular in recent years, cognitive radio networks (CRNs) can be integrated with cloud platform. One of the important issues in CRNs is security. It becomes a problem since CRNs use radio frequencies as a medium for transmitting and CRNs share the same issues with wireless communication systems. Another critical issue in CRNs is performance. Security has adverse effect to performance and there are trade-offs between them. The goal of this paper is to investigate the performance related to security trade-off in CRNs with supporting cloud platforms. Furthermore, Queuing Network Models with preemptive resume and preemptive repeat identical priority are applied in this project to measure the impact of security to performance in CRNs with or without cloud platform. The generalized exponential (GE) type distribution is used to reflect the bursty inter-arrival and service times at the servers. The results show that the best performance is obtained when security is disabled and cloud platform is enabled.

## I. INTRODUCTION

THE exponential growth of wireless devices in the last decade leads to an increasing of spectrum demands [3]. Meanwhile as it has been known radio frequency is a natural resource that cannot be replaced and it has a limitation [4]. Therefore the scarcity of spectrum becomes a problem [8] and radio frequency grows into one of the most priceless resources [9]. However a recent study shows that a great number of licensed spectrums are occasionally operated and it has differences in the spectrum utilization that range from 15% to 85% [7]. It means that most of the allocated frequencies are under-utilized [9]. For example, a study in [10] shows that in the USA, radio frequencies assigned to cellular networks start to increase since morning at the beginning of work times and touch the maximum utilization throughout working hours, but continue unused from midnight until early morning.

Kurniawan D. Irianto is with the Department of Informatics, Islamic University of Indonesia, Yogyakarta, Indonesia (phone: +62 274895287; fax: +62 274-895007; e-mail: k.d.irianto@uii.ac.id).
Demetres D. Kouvatsos is with the School of Electrical Engineering and Computer Science, University of Bradford, Bradford, BD7 1DP UK.

Cognitive radio is a technology in wireless communication which goals to improve the radio frequency utilizations [2]. In terms of users using the frequencies, the Federal Communications Commission (FCC) divided it into two categories; licensed or primary users (PUs) and unlicensed or secondary users (SUs) [5]. In cognitive radio networks, SUs are allowed to use the licensed spectrum temporarily as long as PUs are idle. Thus SUs are expected to have the ability of spectrum sensing and detect any potential activities of PUs [7].

Cloud computing has become popular in the present years [11] but cloud computing is not a new concept at all [12]. The "reusability of IT capabilities" becomes a basic principal of cloud computing inspiration and the difference that cloud computing brings compared to traditional concepts of "grid computing", "distributed computing", "utility computing", or "autonomic computing" is to broaden horizons across organizational boundaries [14]. Now days, cloud computing consists of a set of software, hardware and interfaces that are able to send services based on a demand from users and this model is called pay-as-you-go model. The foremost benefits of cloud computing are availability, scalability and maintenance fees. When CRNs meet cloud computing, CRNs will become more intelligent than before [13].

As it has been known, CRNs have the same issues like wireless networks because of its nature of openness. Attacks and hacking in wireless networks are expected [15]. Wassim et al. [16] divided attacks in CRNs into four main categories: Physical Layer Attacks, Link Layer Attacks that known as MAC Attacks, Network Layer Attacks, and Transport Layer Attacks. Security issues in CRNs keep increasing as long as wireless and mobile devices are produced massively and this is the main reason why security threats in CRNs are become a challenging study in CRNs [15].

In this paper, the trade-off is investigated between performance and security for an admission control server in cognitive radio networks with cloud platforms. It is assumed that there are two kinds of user namely primary users (PUs) and secondary users (SUs). When the security is enabling, how much does the system performance will be affected and when the system is supporting with or without cloud, how much does the performance will be different? To answer the questions of this kind, the queuing network models with multiple classes and one/multiple servers under two different service disciplines of priority queuing, namely Preemptive Resume (PR) and Preemptive Repeat Identical (PRI), are analyzed via Discrete Event Simulation (DES). PR priority queue can represent the CRNs with cloud and PRI priority queue is able to model the CRNs without cloud platform. The security mechanism will be "ON" and "OFF" condition to see the effect of networks. In order to replicate the traffic burstiness of the packet arrival process and the variability of service time at the security server, both interarrival and service time distributions are modeled by the generalized exponential (GE) distribution.

This paper is organized as follows: Section II presents as background work, brief description of cognitive radio networks, cloud computing, queuing network models. Attacks on cognitive radio networks are described in Section III. System design with queuing model applied in this study is presented in Section IV. The simulation results and associated comments are showed in Section V. Conclusions and suggestions for furthering investigation follow in Section VI.

## II. BACKGROUND WORK

### A. Cognitive Radio Networks

Cognitive networks are being studied due to the constraint of spectrum availability [6] and most of the frequency bands for wireless communication purposes have been licensed by the FCC. According to the FCC definition, *a "Cognitive Radio" is a radio that can change its transmitter parameters based on interaction with the environment in which it operates* [5]. From the CR's definition, it can be assumed that cognitive radio has two major characteristics [1]:

1. Cognitive Capability

This means a cognitive user is aware of the change in its environment and has the ability to sense all the surrounding information. It can detect unoccupied spectrums, referred to as *spectrum holes*, and use it temporally. In fact, spectrum holes can be shared with the other cognitive users.

2. Cognitive Re-Configurability

As mentioned before, cognitive radio has a capability of spectrum awareness. This makes a cognitive radio can be programmed dynamically based on the radio information in its environment.

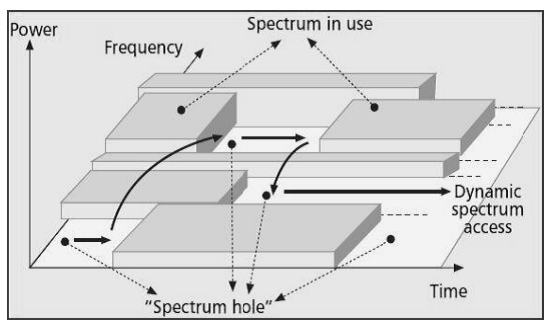

Fig.1 The spectrum hole concept [1]

The idea of cognitive radio (CR) was first presented by Joseph Mitola III [15], [17]-[18]. Ever since researches in CR have grown become popular due to the shortage of radio frequency [18]. CR is a technology that has ability to sense any available spectrum bands and use it in an opportunistic manner at given time [7]. CR allows unlicensed users to, firstly, detect unoccupied frequencies and the presence of licensed users (spectrum sensing), secondly, choose the most excellent available spectrum (spectrum management), thirdly, share the spectrum with the other unlicensed users (spectrum sharing), and lastly, stop using the spectrum whenever the licensed users are present (spectrum mobility) [7].

### B. Cloud Computing

Cloud computing is an evolution of the Internet. The cloud itself refers to a set of hardware, networks, storage, services, and interfaces that allow the delivery of computing as a service [23]. In cloud computing, data and computation are operated somewhere in a "cloud," which is some collection of data centers owned and maintained by a third party [24].

Cloud computing is presented in three forms [23]:

1. Public Cloud

A cloud is made available in a pay-as-you-go manner to the general public. It means that a customer will pay for the service only when he/she use the services.

2. Private Cloud

The cloud infrastructure is operated exclusively for a business or an organization.

3. Hybrid Cloud

A combination of the private and public cloud is called Hybrid Cloud.

In clouds, businesses and users access services based on their requirements from anywhere in the world without regard to where the services are hosted. Many computing service providers, including Microsoft, Yahoo, Google and IBM are rapidly deploying data centers in various locations around the world to deliver Cloud computing services. These data centers host a variety of applications on shared hardware platforms. The applications include serving requests of web applications that only run for a few seconds, large data set processing that run for longer periods of time, distributed databases that need real time response and internet banking that requires security guarantees. The need to manage multiple applications in a data center creates the challenge of on demand resource provisioning and allocation in response to time varying workloads [22].

## III. ATTACKS ON COGNITIVE RADIO NETWORKS

Threat is a constant danger through people, objects, or any resources whereas an attack is an act of or event that exploits the vulnerability. The policies, learning mechanisms, and self-propagation in cognitive radio architecture prevents the threats (cannot escape the threats). In CR, a threat can happen while sensing of information (due to involvement of a malicious user). This information will then feed for learning and decision making. The results produced will lead to inappropriate decisions (unacceptable decisions) due to a malicious user injected the faults [15]. Table I shows the attack types, network layers involved, and reason for attacks.

The attacks on CRNs can be grouped into four major classes: Physical Layer attacks, Link Layer attacks (also known as MAC attacks), Network layer attacks, and Transport Layer attacks [16].

TABLE I
ATTACKS BASED ON LAYERS IN COGNITIVE RADIO

| Attack Type | Network Layer | Reason | Countermeasures |
|---|---|---|---|
| Primary and secondary user Jamming | Physical | Lack of knowledge about location and unclear access rights to cognitive user | Location Consistency Checks. Compare signal strength and noise level |
| Primary signal sensing | Physical | Low level primary signal will be missed | Energy-based sensing. Waveform-based sensing. Cooperative detection of PU |
| Overlapping secondary users | Physical | Location based. Hard to prevent | Use game models and Nash equilibrium techniques to detect transmission power of SUs |
| SUs unauthorizedga in in Bandwidth by pretends as primary user or False feedback | MAC | Malicious SU tweaks withhigher powerbandwidth, and feed falseinformation togain signal | Trust management on secondary users for resource hungry and collaborative trust. Management of systems objective function by controlling the radio parameters |
| Increase interference by malicious node | Network | Compromising with malicious node | Appropriate local spectrum sensing controller. Eliminating internal hidden parasite nodes |
| Ripple effect | Network | False information about spectrum assignment | Continuous trust management process on SUs |
| Key duplication | Transport | Breaks the cipher system | Reinvestigate the protocol activity in the context of sessions. Use secure protocols with robust distribution of key management |
| Jelly fish | Transport | Effect on throughput | Trust of node by verifyingthe packet loss |

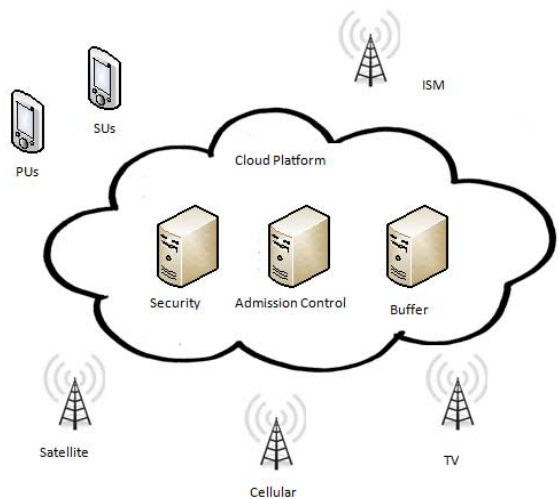

Fig. 2 Illustration of the CRNs with cloud platforms

## IV. SYSTEM DESIGN WITH QUEUING MODEL

This project is an investigation of performance vs. security in cognitive radio network and supporting cloud platform based on the algorithm proposed by Rawat et al. [20]. However a new model of queuing system is developed to investigate the performance and security. Furthermore, a queuing network model is built in order to measure the performance of the system and explore the impact of the security on the performance.

As shown in Fig. 2, it is assumed that there are 4 different networks, namely Industry Scientific and Microwave (ISM), Satellite, Cellular and TV network, reported its status of idle channels to the cloud. The SUs will connect to the cloud whenever it is looking for an idle channel and access the unoccupied spectrum in condition that PUs are not using the spectrum. If the SUs detect PUs are accessing the spectrum, SUs will stop transmitting and vacate the channel. SUs may be rejected by Admission Control if its geo-location, data rate demanded and payment offered are not matching with the available spectrum requirements. Fig. 6 presents the illustration of the model in this project. In the cloud, there are three servers such as security, admission control and buffer server.

The security server is to ensure the trustworthiness of PUs. It can be grouped into two techniques; securities based on location and transmit power level technique [21]. The buffer server is used to store the rest of transmission of SUs. Whenever there is an available channel SUs will resume the transmission.

### A. Queuing Network Model (QNM)

Fig. 3 represents the queuing network model of this project. PUs and SUs have to go through three steps in order to access the channels. The first step is security. The behavior of PUs will be checked by security server. If PUs are detected as malicious users, security server will drop its request. However if PUs are trustworthy users, security server will allow its request to go forward to the next server.

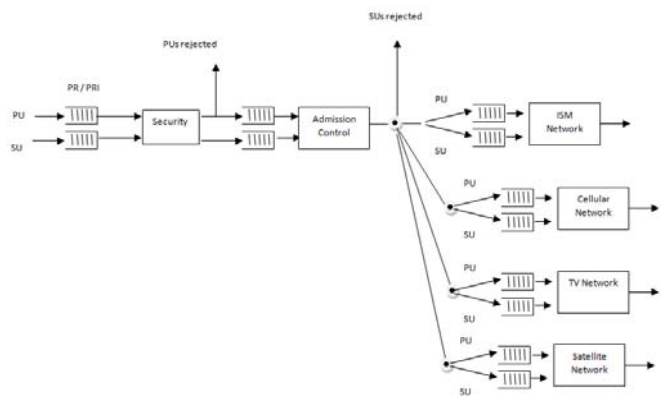

Fig. 3 Queuing network model for this project

The second step is Admission Control (AD). AD will check the requests of PUs and SUs and forward its request to the available channel. SU request may be drop if its request of date rate, payment offered and geo-location does not match with the available channel requirements. AD will forward SU request to access the unused channel. The final step is to access the available channel. SUs will use the available channel as long as PUs are idle. Whenever PUs are detected, SUs will vacate the channel and go back to the front of SU's queue. If there is any unoccupied channel, SUs will resume the transmission. The GE-type distribution is used. The performance of the system will be observed whenever the security node is ON or OFF and with or without cloud.

### B. QNM Implementation

A priority queuing system can represent the performance of CRN either with or without cloud platform. In cloud platform, we make an assumption there is a buffer server for storing the SUs transmission when SUs are interrupted by PUs. Therefore CRN with cloud platform can be modeled by *pre-emptive resume (PR) priority queue*. On the other hand, *pre-emptive repeat identical (PRI) priority queue* is able to model the CRN without cloud platform because SUs will start over its transmission when it gets interrupted by PUs.

## V. SIMULATION AND RESULT

### A. Simulation

Discrete-event simulation [19] was implemented using a java package to simulate GE/GE/C/N/PR for the system with cloud platform and GE/GE/C/N/PRI for the system without cloud in terms of mean response time, mean queue length, utilization and packet loss. The program was run up to 20 independent times using GE-type distribution. Table II shows the simulation schemes in this study.

TABLE II
SIMULATION SCHEMES

| Scheme | Criteria | Description |
|---|---|---|
| A | PQ = PR, PRI; SEC = ON, OFF; c = 1; PU = 3; SU = 1-6; N = 20; SCV = 1; GE/GE/1; MU = 13. | Assessing the effect of ON/OFF of security and cloud platform |
| B | PQ = PR; SEC = ON, OFF; c = 1; PU = 1, 3, 5; SU = 1-6; N = 20; SCV = 1; GE/GE/1; MU = 13. | Measuring the effect of the PU requests |
| C | PQ = PR; SEC = ON, OFF; c = 1; PU = 3; SU = 1-6; N = 20; SCV = 4, 8, 10; GE/GE/1; MU = 13. | Performing the effect of SCV |
| D | PQ = PR; SEC = ON, OFF; c = 1, 3; PU = 3; SU = 1-6; N = 20; SCV = 1; GE/GE/1; MU = 13. | Assessing the effect of the number of servers |

PQ = Priority Queue, PR= Preemptive Resume, PRI= Preemptive Resume Identical, Sec= Security, PU= Primary User, SU= Secondary User, c= Number of Server, SCV= Number of Squared Coefficient of Variation, N= Capacity of Queue, MU= Mean Service Rate, GE= Generalized Exponential Distribution.

### B. Results

Figs. 4-19 show the simulation results of Table II.

1. Assessing the Effect of On/Off of Security and Cloud Platform

Figs. 4 to 7 illustrate the different of the system when security is enabled /disabled and the cloud platform is ON/OFF. Fig. 4 depicts the mean total waiting time in queue of scenario A. It can be seen that the system with security off and cloud platform has the lowest value of waiting time which means it is the best performance. Meanwhile the system with security on and without cloud has the worst performance. The waiting time in security off is better than security on. It is because when security is on, the system needs more time in processing the SU requests.

It is shown in Fig. 5 that the longest of SUs spent in the system is when the system with security on and without cloud platform. As expected, security has impact on the performance and without cloud platform. SUs are not able to store its transmission whenever it is interrupted by PUs and have to start over the transmission from the beginning. However, the mean response time in the system with security off and cloud platform is the lowest and has the best mean response time.

Fig. 6 presents the mean queue length of scenario A. The system with cloud platform introduces the best performance. Meanwhile, the graph in Fig. 7 illustrates total packet loss in scenario A. The system without cloud and security on has the highest value of packet loss which means it is the worst performance. There is almost no packet loss in the system with cloud and security off.

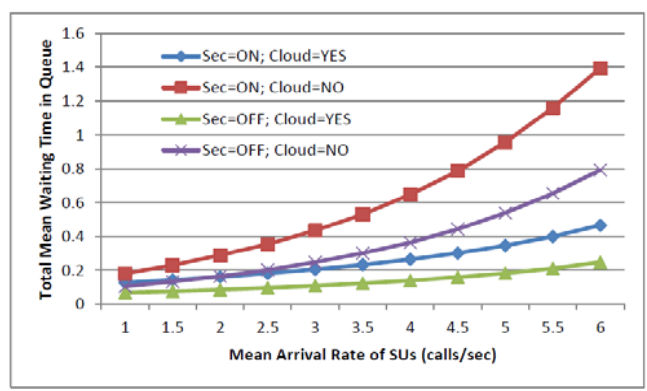

Fig. 4 Total mean waiting time in scheme A

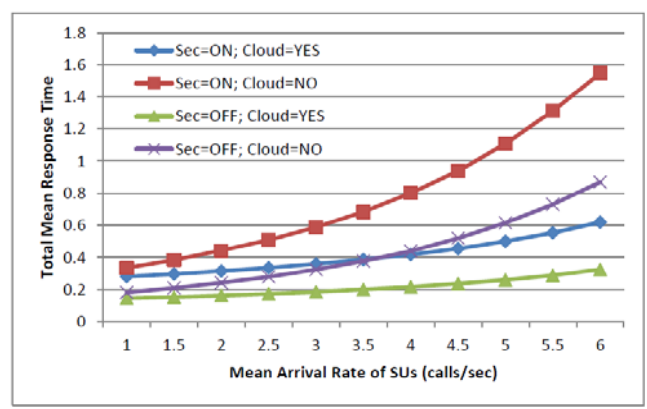

Fig. 5 Total mean response time in scheme A

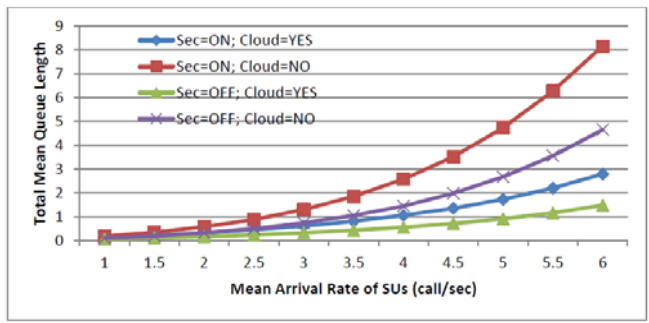

Fig. 6 Total mean queue length in scheme A

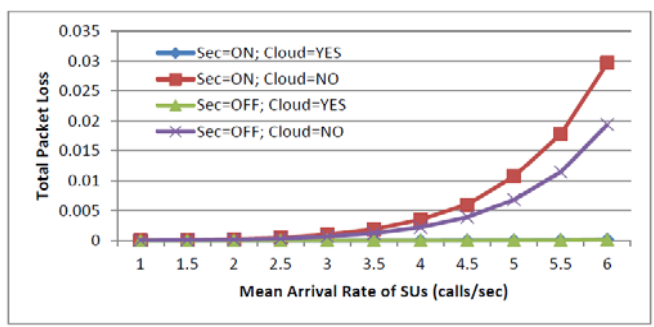

Fig. 7 Total packet loss in scheme A

2. Measuring the Effect of the PU Requests

It shows in Fig. 8 that when PU is 1 call/sec, mean waiting time is smaller than when PU is 3 or 5 calls/sec. The best waiting time is shown when the security is off and PU is 1 call/sec.

In Fig. 9, it presents the average response time when security is on/off and PU is set to 1, 3, 5 calls/sec. The graph shows that the best performance in terms of response time is when security is off, cloud is enabled and PU is 1. In contrast, the worst response time occurs when the security is ON, cloud is YES, and PU is 5.

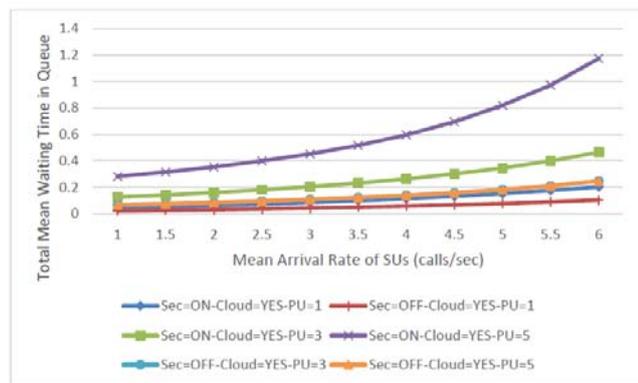

Fig. 8 Total mean waiting time in scheme B

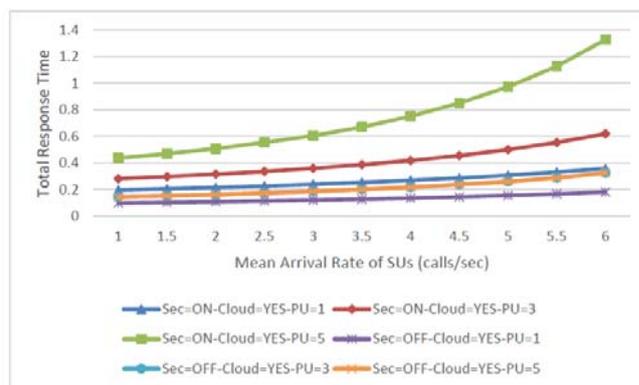

Fig. 9 Total mean response time in scheme B

The security and arrival rate of PUs have influence on the performance. The bigger value of PUs, the longer mean queue length of SUs is. From Fig. 10, it seen that mean queue length has the smallest value when security is OFF, cloud is YES, and PU is 1. The mean queue length is better if the security is OFF than security is ON.

It shows in Fig. 11 that when security is ON and PU value is 5, total packet loss is the worst and it reaches almost 0.009 when mean arrival of SUs is 6 calls/sec. The best packet loss is when security is OFF, cloud is YES, and PU is 1.

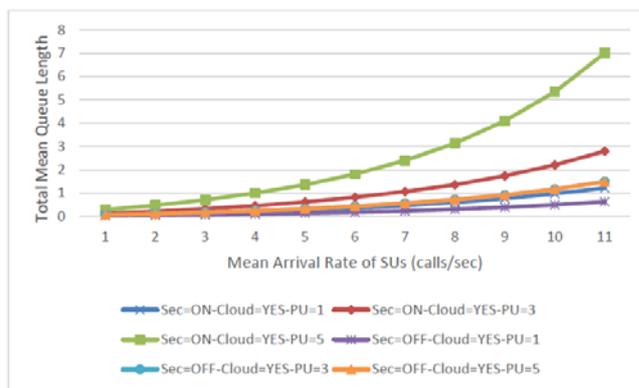

Fig. 10 Total mean queue length in scheme B

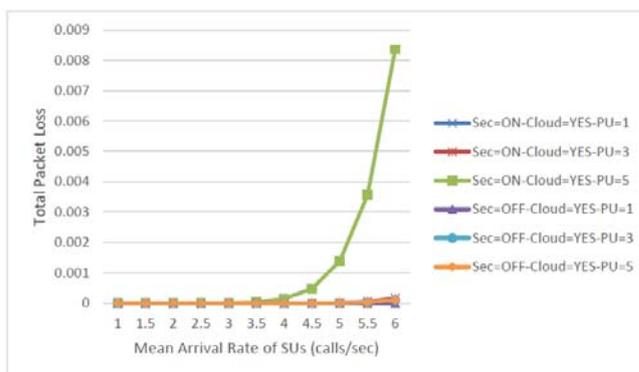

Fig. 11 Total packet loss in scheme B

3. Performing the Effect of SCV

The average waiting time in queue presented in Fig. 12 has the best value when the security is OFF, cloud is YES and SCV is 4. On the other hand, the waiting time takes a longer time when the security is ON and the value of SCV is high. Figs. 13-15 illustrate that the performance in terms of response time, mean queue length, and packet loss is better if the SCV is small and security is OFF. The SCV presents the traffic burstiness in the networks and affects the performance.

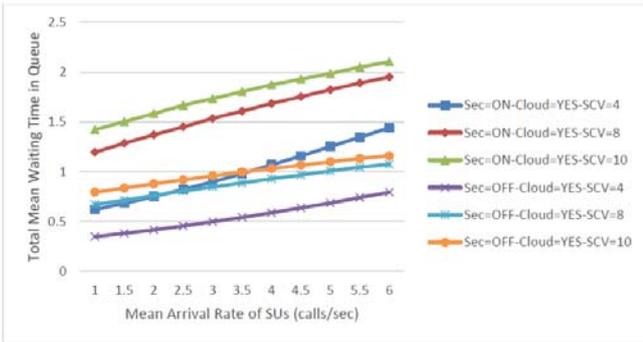

Fig. 12 Total mean waiting time in scheme C

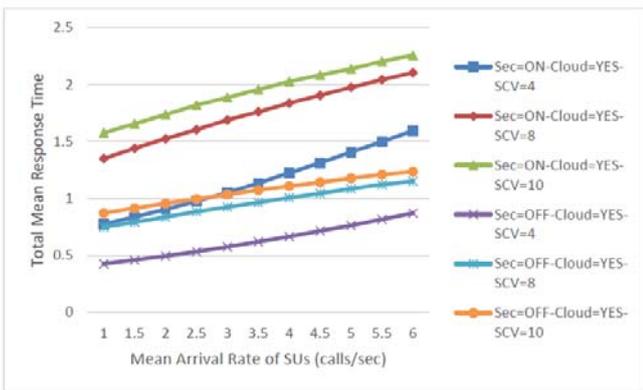

Fig. 13 Total mean response time in scheme C

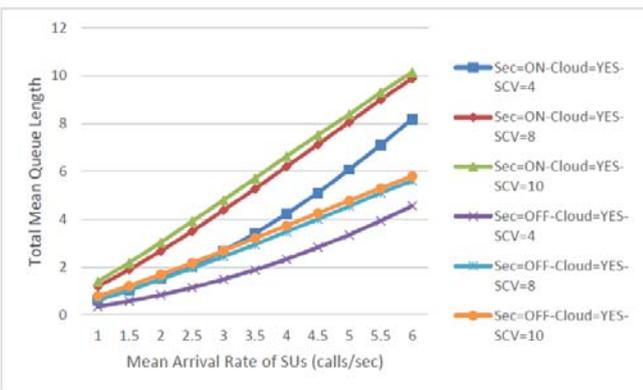

Fig. 14 Total mean queue length in scheme C

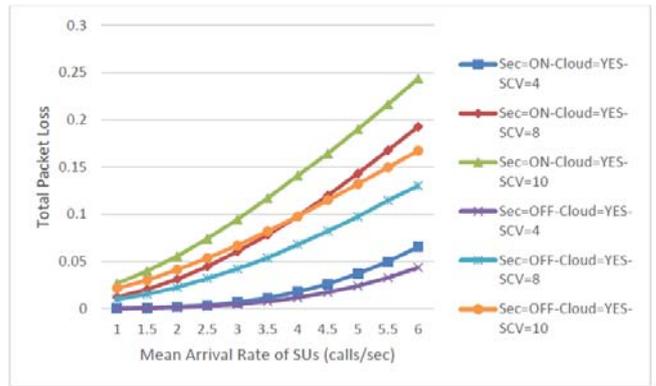

Fig. 15 Total packet loss in scheme C

4. Assessing the Effect of the Number of Servers

The average waiting time and response time are presented in Figs. 16 and 17. It is noticeable that there is a different of performance in single and multiple servers which multiple servershave better performance. Moreover, the performance also depends on the security and cloud platform. The best performance is obtained when security is off, cloud is enabled and server is 3. In contrast, the worst performance occurs when security is on, cloud is disabled and server is 1.

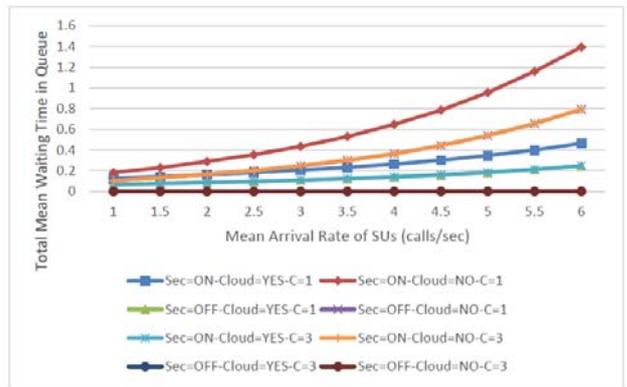

Fig. 16 Total mean waiting time in scheme D

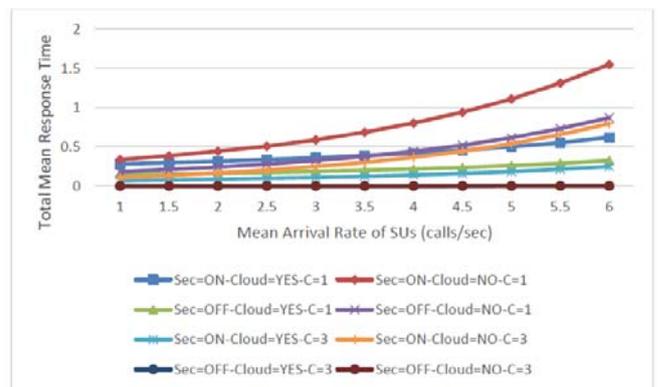

Fig. 17 Total mean response time in scheme D

The values of mean queue length and packet loss are seen in Figs. 18 and 19. It is quite clear that the packet loss and mean queue length have the best performance when security off, cloud is YES and server is 3. It deduces that multiple server is better than single server.

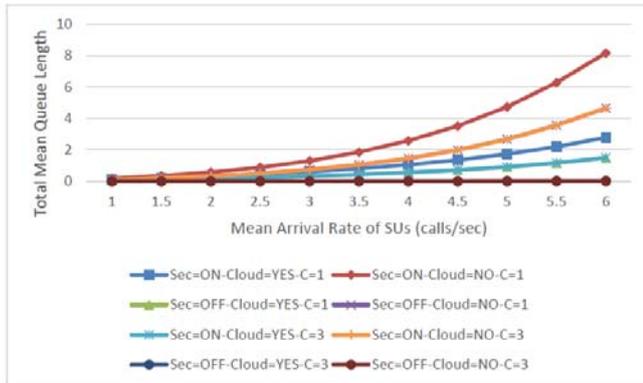

Fig. 18 Total mean queue length in scheme D

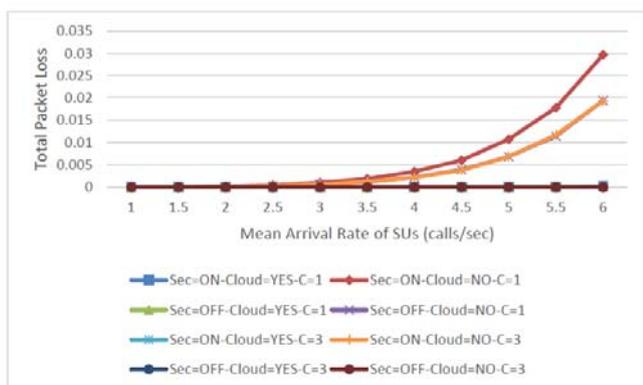

Fig. 19 Total packet loss in scheme D

## VI. Conclusion

From scheme A, B, C, and D, it can be concluded that cognitive radio network with cloud platform is better than without cloud platform and security has adverse impact on the performance. The best performance will be obtained if the security is off and with cloud platform. However, without security, it will make the system is vulnerable. The presence of PUs will degrade the performance of SUs and multiple servers have better performance than single server. Moreover, the traffic burstiness has impact on the performance.

For the future work, this research can be extended by considering the PBS (partial buffer sharing) and CBS (complete buffer sharing) policies to evaluate the performance in different situations. In addition, the project could be enhanced to cognitive radio ad-hoc network.